\RequirePackage{silence}
\documentclass{aastex631}
\usepackage{graphicx}
\usepackage{xcolor}
\usepackage{diagbox}
\usepackage{appendix}
\usepackage{amsmath}
\usepackage{comment}

\usepackage{chngcntr}
\bibpunct{(}{)}{;}{a}{,}{,}

\begin{document}

\title{TitanMPAS - a new global and regional Titan atmospheric circulation model. Part I: Global model results}
\author{Yuan Lian}
\affiliation{\it{Aeolis Research, Chandler, AZ}}

\author{Mark I. Richardson}
\affiliation{\it{Aeolis Research, Chandler, AZ}}

\author{Claire E. Newman}
\affiliation{\it{Aeolis Research, Chandler, AZ}}

\author[0000-0001-8233-2436]{Martin A. Cordiner}
\affiliation{Astrochemistry Laboratory, NASA Goddard Space Flight Center, 8800 Greenbelt Road, Greenbelt, MD 20771, USA.}
\affiliation{Department of Physics, Catholic University of America, Washington, DC 20064, USA.}

\section{Abstract}

We developed TitanMPAS, a new Titan global and regional climate model, to simulate atmospheric dynamics and methane cycles. It integrates the non-hydrostatic MPAS dynamical core and TitanWRF physics parameterizations, with corrections to relax the thin-atmosphere assumptions on the gravity and curvature terms. TitanMPAS improves upon TitanWRF by incorporating a better moist convection parameterization and the observed surface properties. Long-term simulations ($>$100 Titan years) with a high resolution of $1^\circ$ and the realistic topography show that TitanMPAS is able to produce the stratospheric superrotation with typical equatorial wind speed over $\rm 120\,ms^{-1}$ and the maximum zonal wind speed exceeding $\rm 170\,ms^{-1}$ . Besides the good match of temperature profiles to the Cassini CIRS observations, the simulated wind, temperature and methane abundance in the troposphere match well with Huygens probe in-situ measurements. TitanMPAS also captures the major features of seasonal methane precipitation distribution patterns inferred from the observed cloud activities, as well as the vertical and latitudinal variations of methane abundance in the reanalysis of Cassini VIMS data. Resolution sensitivity studies show that higher resolution produces faster stratospheric superrotation, as well as stronger near-surface wind due to better resolved terrain variations. We further demonstrate that, despite the weak solar insolation, the near-surface temperature, pressure, wind magnitude and methane abundance all exhibit diurnal variations. Lastly, we show potential impacts of topography on methane transport and eddy momentum flux in both troposphere and stratosphere, which should be investigated in a future study. 

\section{Introduction}

Titan, the largest moon of Saturn, has a thick, nitrogen rich atmosphere that assimilates an Earth-like environment where an active hydrological cycle of methane produces rain and lake formations. Despite its slow rotation rate, the Titan's stratosphere exhibits strong prograde wind with speed exceeding $\rm 180 \,ms^{-1}$ \citep{Achterberg2011,Achterberg2023}, which is a superrotating atmosphere similar to that of Venus. Conversely, the tropospheric wind is quite weak with typical wind speed of $\rm \sim 40 \, ms^{-1}$ near the tropopause and reduces to a merely a few meters per second in the lower 20 km above the surface \citep{Flasar2005,Bird2005}. Unlike Earth, Titan has a relatively flat topography that varies a few hundred meters relative to its geoid \citep{Lorenz2013}. The Titan's surface appears to be darker overall while the thermal inertia can vary by an order of magnitude, particularly at the low latitudes where the surface compositions are more complicated and at the high latitudes where lakes form \citep{Solomonidou2018,MacKenzie2019}.

The dynamic structures of Titan's lower atmospheres (troposphere and stratosphere) are primarily studied by numerical models, ranging from early 2D models to the more complex 3D global circulation models (GCM) with physics parameterizations that represent the physics processes including atmospheric radiative transfer, planetary boundary layer mixing, methane moist convection and surface/subsurface models that handle the surface torque and surface-atmosphere heat and tracer exchanges. The quality of these models is quantified by the comparison between model predictions and the limited observations of temperature, wind, clouds and lake distributions, {\it i.e.}, measurements by Cassini Composite Infrared Spectrometer (CIRS) and Huygens probe, as well as the ground based Atacama Large Millimeter/sub-millimeter Array (ALMA) wind measurements \citep{Flasar2005,Achterberg2011,Rannou2021,Cordiner2026,Turtle2011,Nixon2025,DeBatzDeTrenquelleon2026}.

The first fully three-dimensional simulation of Titan's global circulation was performed by \citet{Hourdin1995}, utilizing a GCM derived from the Laboratoire de Météorologie Dynamique (LMD) terrestrial model. This early model demonstrated that stratospheric superrotation could be initiated and maintained from a state of rest through the upward transport of angular momentum by mean meridional circulations and barotropic waves. Based on the original LMD framework, the Institute Pierre Simon Laplace (IPSL) Titan GCM extended model top up to 500 km, where the upper stratospheric dynamics and photochemistry could be studied \citep{Lebonnois2012a}. The IPSL GCM later evolved to the Titan Planetary Climate Model (Titan PCM), which coupled a complete atmospheric physics package with active methane hydrology, haze microphysics, and cloud nucleation schemes \citep{DeBatzDeTrenquelleon2025}. Another early Titan GCM,  the University of Cologne (Köln) Titan GCM, developed by \citet{Tokano1999}, adopted the quasi-hydrostatic grid-point dynamical core of the ARIES/GEOS GCM to study tropospheric wind fields, methane condensation, and the effects of Saturn’s gravitational tides.

TitanWRF was among the first that produced stratospheric superrotation with speed comparable to that derived from the CIRS temperature measurements. This Titan GCM is based on the  Weather Research and Forecasting (WRF) finite difference dynamical core, and it is a component of planetWRF \citep{Richardson2007}. TitanWRF mainly simulates the dry atmosphere without topography \citep{Newman2011}, but the seasonal methane hydrological cycles and their impact on the stratospheric winds were examined using a simplified large-scale condensation scheme \citep{Newman2016}. A more modern Titan GCM, Titan Atmospheric Model (TAM), was developed by \citet{Lora2015}, utilizing the Geophysical Fluid Dynamics Laboratory (GFDL) Flexible Modeling System (FMS) spectral dynamical core, combining non-gray radiative transfer with a simplified Betts-Miller moist convection scheme to explore the seasonal methane hydrologic cycle. Additionally, specialized adaptations such as the Titan Community Atmosphere Model (CAM-Titan)—a derivative of NCAR's CAM 3.0 coupled with the CARMA aerosol code—were employed to quantify the impact of regional topography, surface roughness, and zonal torques on dune-forming surface wind fields \citep{Larson2015}. 

All these Titan GCMs achieve various degrees of success.  IPSL/PCM, TAM, and TitanWRF are the notable GCMs that are able to produce stronger stratospheric superrotation, with the wind speed in TitanWRF being closer to those retrieved from the Cassini CIRS observations. Other GCMs, with particular versions of IPSL/PCM and TAM, mainly focuses on the tropospheric dynamics and hydrological cycles. Among these models, TAM produces better match to the observed wind, temperature and methane profiles in the lower 40 km by Huygens probe. Despite the success, none of the published Titan GCMs is able to produce strong stratospheric superrotation with the presence of topography. Moreover, all these GCMs utilize relatively coarse grid (typically $4^\circ-5^\circ$ globally). This resolution is sufficient to resolve large-scale dynamics since the Rossby deformation radius in Titan's atmosphere is large (due to large static stability and small rotation rate), but it may not be able to capture the detailed near-surface wind variation and eddies with orographic origin that are directly impacted by the resolution of topography.

Here we present a new Titan GCM based on the Model for Prediction Across Scales (MPAS), TitanMPAS. This model is intended to simulate the tropospheric and stratospheric circulations simultaneously with the presence of topography and active methane cycle. Previous GCMs assumed that the stratospheric circulation has little impact on the tropospheric dynamics as these are two distinct dynamic regimes ({\it i.e.}, $R_o<<1$ in the troposphere and $R_o>>1$ in the stratosphere with a near-zero wind layer sitting at 70-80 km), thus relaxing the requirement of stratospheric superrotation when simulating physical and dynamical processes in the troposphere. However, this approach requires a forced or an ad-hoc upper boundary condition that may not be consistent with the stratospheric circulation patterns. Furthermore, such assumption has not been thoroughly examined and validated in a single, self-consistent model that captures the characteristics of tropospheric and stratospheric circulations simultaneously with the presence of topography, which generates orographic waves that may interact with the wind-minimum layer at 75 km (if it acted as a critical layer) or propagate to the upper stratosphere where they exert drag upon reaching saturation. Additionally, the existing Titan GCMs utilize either spherical-polar grids with finite difference/finite volume dynamical or spectral dynamical core, with the former being sensitive to the polar convergence problem and the latter being inefficient to scale up the computation speed with high-spatial resolutions. A GCM with uniform meshes and efficient scalability is highly desired for Titan that requires long integration time ($>$ 70 Titan years) due to its large self-rotation and orbital period with very weak insolation.

TitanMPAS GCM contains two components: global and regional models, which will be presented in a series of publications. In this study, we only present the global model results as Part I with limited diagnostics. The regional mesoscale simulations will be presented in Part II. The detailed model diagnostics including the wind driven mechanisms due to wave-mean interactions, particularly how the orographic waves can impact the stratospheric superrotation ({\it e.g.,} \citet{Flowers2024}), as well as the tracer cycles due to moist convection and other dynamic processes (including wind and thermal interactions) will be presented in Part III. This paper is organized as the following: Section~\ref{model_description} describe TitanMPAS model, Section~\ref{model_configuration} describe model configurations, Section~\ref{model_results} presents simulation results, and Section~\ref{conclusion_discussion} concludes with open questions. 

\section{TitanMPAS description}
\label{model_description}

TitanMPAS is developed based on the MPAS dynamical core. MPAS solves the fully compressible and non-hydrostatic governing equations of continuity, moment and heat in a three-dimensional atmosphere \citep{Skamarock2012}. TitanMPAS is a newly added component to planetMPAS, which was initially developed to simulate the Martian and Jovian atmospheres \citep{Lian2023}. Here we focuses on the modifications made to the dynamical core and physics parameterizations that are specifically related to Titan.

TitanMPAS shares key physics parameterizations with TitanWRF, including the two-stream radiative transfer mode with seasonal haze distribution, the planetary boundary layer (PBL) scheme that mixes heat, momentum, and tracers in convectively unstable atmosphere, surface and subsurface scheme that exchanges heat between surface/subsurface soil and the near-surface atmosphere \citep{Newman2011}. In the current version of TitanMPAS, the radiative transfer model (RTM) has been streamlined to utilize an unified k-distribution model (KDM) originally developed for Martian atmosphere \citep{Mischna2012}, where methane abundance is now predicted from methane hydrological cycle. Simulations with the KDM model produce better thermal profiles in the troposphere, notably it correctly produces tropopause near 0.1 bars with a temperature of about 71 K, whereas the RTM in Newman et al. 2011 produced tropopause near 0.2 bars with a much cooler temperature at about 67 K. The methane moist convection is represented by a Betts-Miller type bulk convective adjustment model, as opposed to the large-scale condensation model in \citet{Newman2016}. This moist convection model is similar to that in \citet{Lora2015}, but with options for implementing either the simplified Betts-Miller scheme \citep{Frierson2007} or a more classic Betts-Miller scheme \citep{betts1993}. The surface methane sources follow the observed lake distributions \citep{Horvath2016} with the assumption of infinite depth. The molecular weight of methane is also factored into the dynamical core and physics parameterizations throughout the model.

\begin{figure}
\centering
\includegraphics[width=1\linewidth, angle=90]{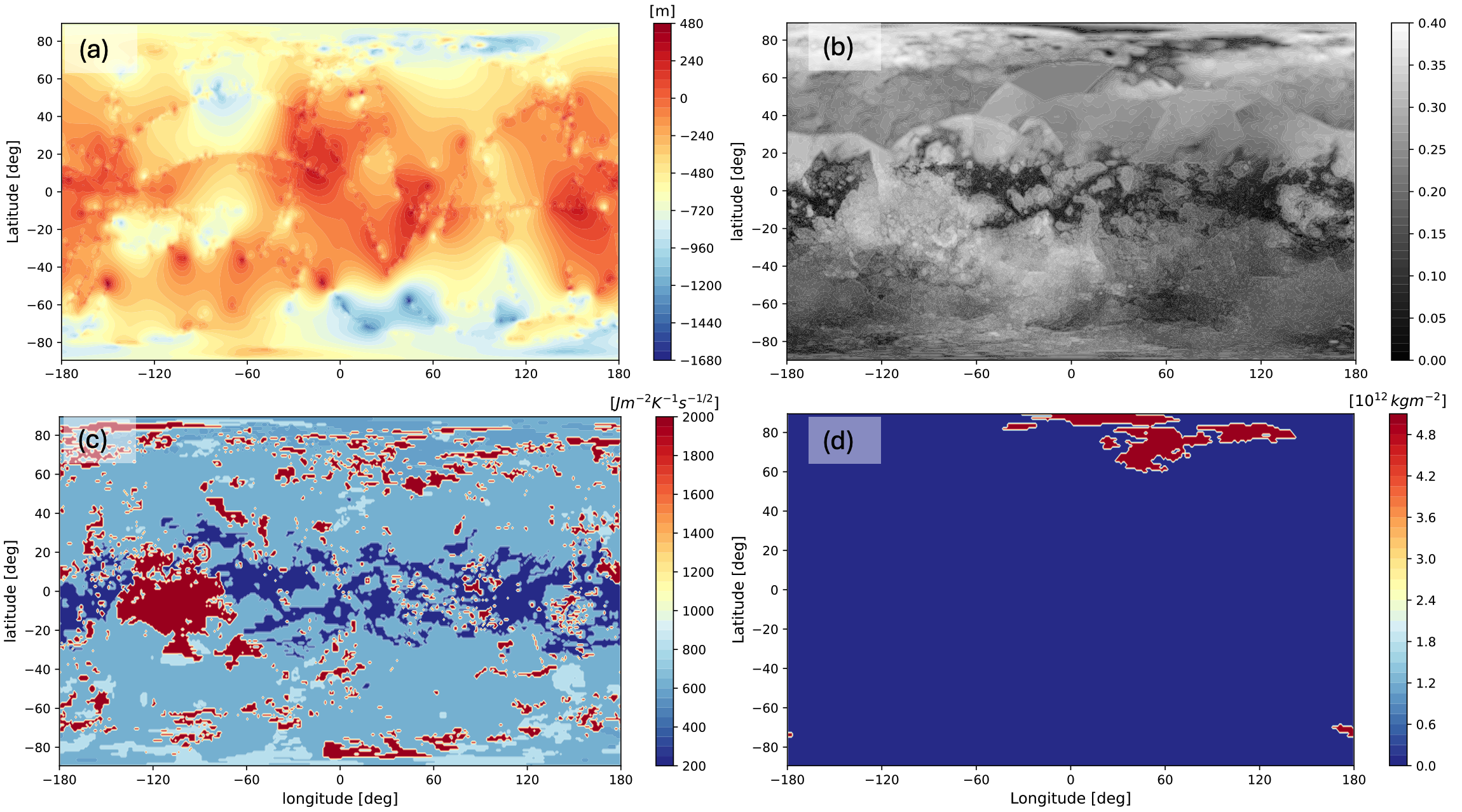}
\caption{\label{fig:topo}Titan surface properties: topography (a), albedo (b), thermal inertia (c) and lake distribution (d). All three maps use $180^\circ$ W to $180^\circ$ E convention.  The Huygens lander location is near $167^\circ$ E and $10^\circ$ S in the topography map.}
\end{figure}

Topography is now a default feature for TitanMPAS (Fig.~\ref{fig:topo}a), including its radiative effect and related orographic wave parameterizations \citep{Lian2023}. The topography is based on the $1^\circ\times 1^\circ$ map in \citet{Lorenz2013}, and it can be up-scaled to embed small features such as the Selk crater in regional-domain simulations. Here we shift the map from \citet{Lorenz2013} by $180^\circ$ E thus placing $0^\circ$ longitude to be the sub-Saturnian point at the northern spring equinox ($L_s=0^\circ$), which is the default initial time of all Titan simulations started from an atmosphere at rest. One benefit of this map definition is that it is consistent with the starting point of Titan in the star-fixed inertial frame, if we ever need to investigate the Titan's stratospheric tilt \citep{Wright2025}. In this map, the longitude for Huygens probe is at $167^\circ$ E.  TitanMPAS also uses surface albedo map (with 4-km resolution map scaled by a global mean of 0.2, Fig.~\ref{fig:topo}b) and thermal inertia map ($1^\circ\times 1^\circ$ with still lakes, Fig.~\ref{fig:topo}c) inferred from the Cassini observations \citep{Solomonidou2018,Lopes2019,MacKenzie2019}. These maps all use the same west-east longitude definition as the topography map for consistency. 

Additional modifications made to TitanMPAS is the altitude dependence of gravity and several specific acceleration terms in the momentum equations. Traditionally the atmosphere is considered to be thin compared to the radii of the terrestrial planets such as Earth and Mars, {\it i.e.}, $z<<a$, where $z$ is the altitude above the surface and $a$ is the planet radius. This implies that gravity is assumed to be constant throughout the atmosphere. For Titan, such assumption may lead to large errors when estimating hydrostatic balance and other planet geometry related acceleration terms due to Titan's small radius and thick atmosphere. For example, near the upper stratosphere, the gravity is reduced by more than $25\%$ compared to the value at the surface. We have made corrections to the gravity and curvature terms in both the dynamical core and physics parameterizations (see Appendix). 

\section{Model configuration}
\label{model_configuration}
The nominal grid resolution for global TitanMPAS is $1^\circ$, which quadruples the nominal resolutions of other popular Titan GCMs including TitanWRF. Sensitivity tests with coarser grid resolutions $2^\circ$ and $4^\circ$ show that the seasonal patterns of large-scale wind and temperature fields are insensitive to the change of grid resolution, as expected from previous studies where the dominant wave modes were identified as planetary-scale waves with typical zonal wavenumber of 1 or 2 \citep{Lewis2023TAMWaves,Lian2025}. However, as we will demonstrate in this study, the resolution can influence the overall wind magnitude, particular the near-surface wind that depends on how well the topographic features are resolved. Moreover, while it is computationally more efficient to run higher-resolution simulations using the lower-resolution restart files instead of starting from the rest (e.g., running $2^\circ$ from $4^\circ$ restart files), the higher-resolution model still would require at least a few Titan years to re-establish quasi-equilibrium state (measured by superrotation indices, see \citet{Newman2011,Read2018}).

TitanMPAS utilizes a terrain-following height coordinate. Vertically there are 60 layers extending from the surface to 550 km. The grid is denser near the surface with the highest vertical resolution of $\sim 100-300$ m and gradually increases to 10-15 km above 150 km. The choice of this vertical grid is based on the requirement of resolving typical vertical wavelengths in the stratosphere ($\sim$ 50-100 km depending on the seasons) and the computational efficiency.  For mesoscale simulations (to be presented in Part II), the horizontal and vertical resolutions vary with the locations of interest. Using the Selk crater where Dragonfly is expected to land as an example, we choose a typical resolution on order of 1 km in order to resolve the narrow rim of the crater (about tens of kilometers) to better represent the slope flow. The vertical resolution is also higher accordingly, with the finest resolution of $\sim 1$ m in the lowest model level.

TitanMPAS is run from an atmosphere at rest, initially with a horizontal grid resolution of $4^\circ$. It takes about 70 Titan years for the atmosphere to spin up to a quasi-steady state.  The model results are then interpolated to generate restart files at higher resolutions at $2^\circ$ and $1^\circ$. For the nominal resolution $1^\circ$, we continue to run the model until it reaches more than 100 Titan years. The results then can be used as initial and temporally-forced boundary conditions for the regional-domain simulations, as the high-resolution results can reduce the jittering around the boundaries of regional domain,  {\it e.g.}, a $8^\circ\times 8^\circ$ (equivalent to $\rm 360 \,km\times 360\,km$) domain to resolve Selk crater and its surroundings.

The uniformity of TitanMPAS mesh allow large time steps. For $4^\circ$ resolution, the typical time step is 75 Titan seconds. As the model resolution increases to  $1^\circ$, the time step reduces to 15 Titan seconds, which is comparable to the typical time step used by TitanWRF in its   $5^\circ$ simulations. To reduce the computational cost, we update the radiative heating/cooling rates every 30 Titan minutes, consistent to the value used in TitanWRF. For regional domain simulations, the time step is much shorter to maintain numerical stability. The radiative heating/cooling rates are also updated with a higher frequency, {\it i.e.,} 5 Titan minutes to capture the local radiative effect.

\section{Results}
\label{model_results}

\begin{figure}
\centering
\includegraphics[width=0.6\linewidth]{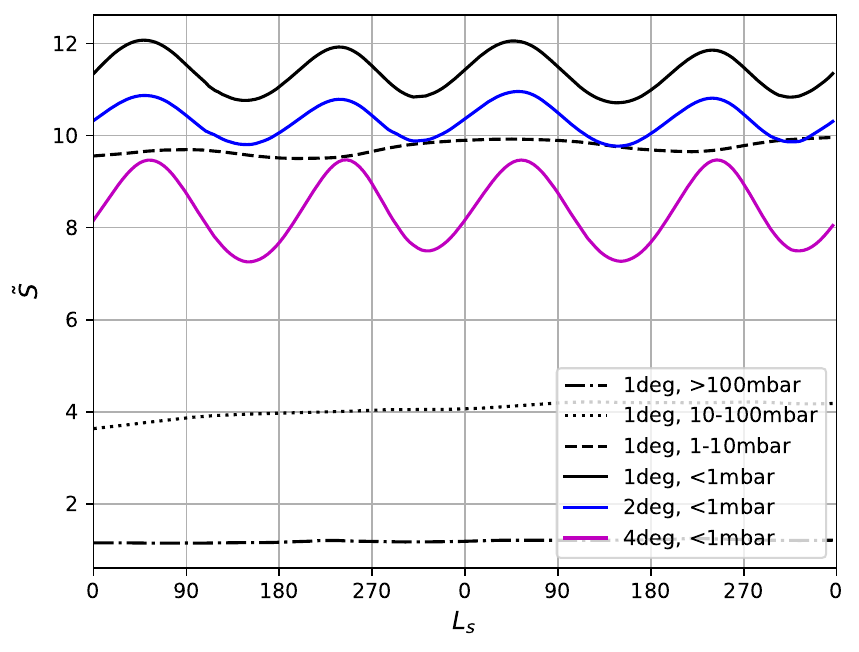}
\caption{\label{fig:sidx_comp}Superrotation indices for simulations with $1^\circ$, $2^\circ$ and $4^\circ$ resolutions. All three cases ran more than 100 Titan years.}
\end{figure}

\begin{figure}
\centering
\includegraphics[width=0.6\linewidth]{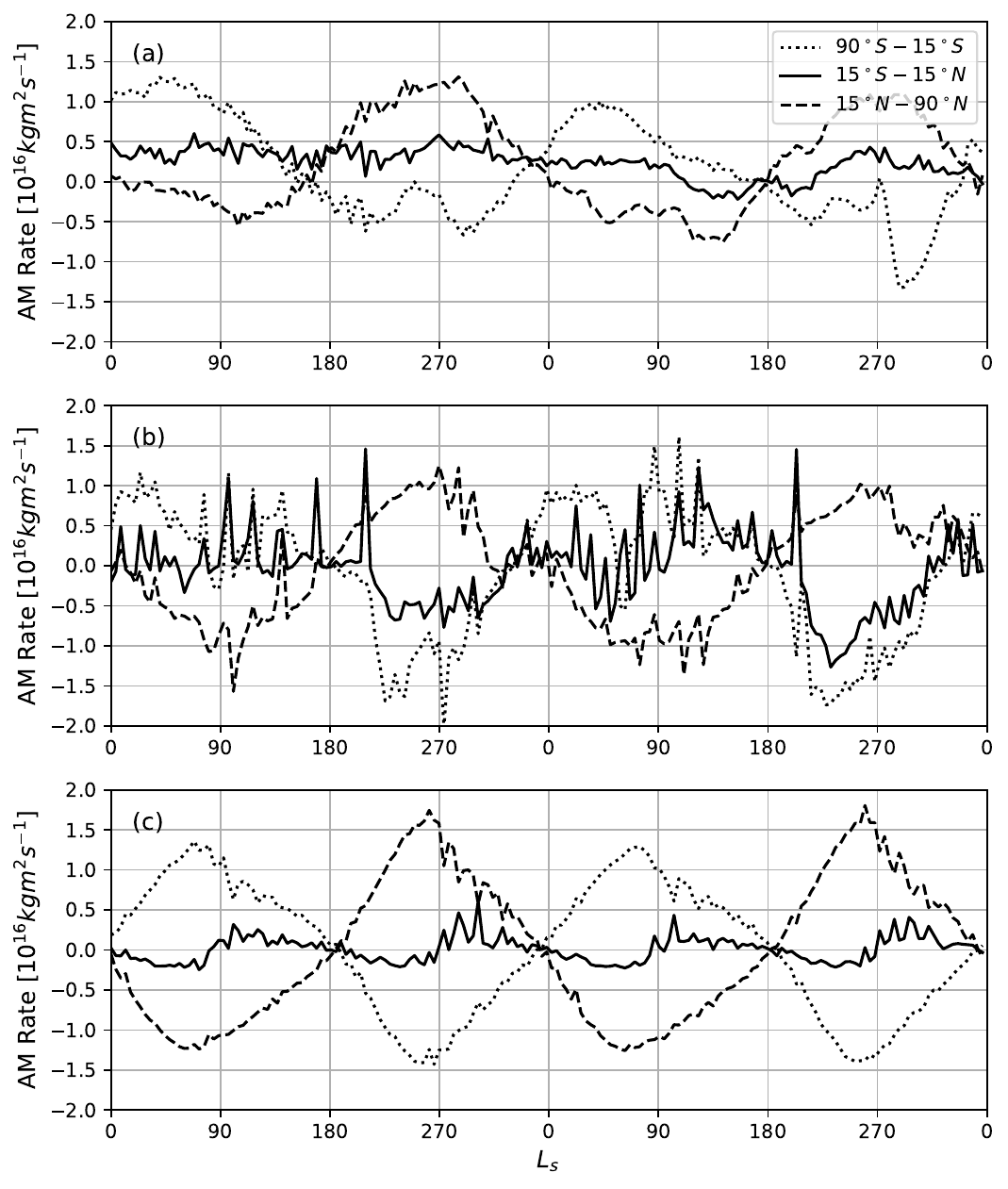}
\caption{\label{fig:am_comp} AM change rate for simulations with $1^\circ$ (a), $2^\circ$ (b) and $4^\circ$ (c) resolutions. The dotted, solid and dashed lines show the AM rate in the southern hemisphere, equatorial region and northern hemisphere.}
\end{figure}

TitanMPAS achieves quasi-steady states after about 100 Titan years for all resolutions tested. Figures~\ref{fig:sidx_comp} and \ref{fig:am_comp} show the superrotation indices and angular momentum change rates in different regions of the atmosphere as a function of Titan years. These quantities are defined slightly different from those in TitanWRF \citep{Newman2011} because TitanMPAS reproduces the observed tropopause near 100 mbars whereas TitanWRF predicted a lower tropopause near 200 mbars. Therefore, for the vertical layers characterizing the dynamic regimes, we define troposphere, lower stratosphere, middle stratosphere and upper stratosphere with pressure ranges $\rm p>100\,mbar$, $\rm 100\,mbar>p>10\,mbar$,  $\rm 10\,mbar>p>1\,mbar$ and $\rm p< 1\,mbar$ respectively. The cutoff latitudes of the equatorial region is set to $\pm 15^\circ$ due to weaker superrotation compared to that in TitanWRF. A more rigorous choice may be obtained via the equatorial Rossby deformation radii ({\it e.g}, \citet{Lewis2023TAMWaves}).  Only 2-year simulation results are shown for comparison purpose among various resolutions. Overall, the long-term variation of the superrotation indices including the spin-up stage (not shown) is similar to those in TitanWRF \citep{Newman2011}.

The superrotation index $\tilde{S}$, defined as the ratio between the absolute angular momentum (AAM) of an atmospheric layer and that of the same layer at rest, is a measurement of overall wind strength, {\it e.g.}, $\tilde{S}=1$ if there is no wind relative to the ground. $\tilde{S}$ remains nearly constant when the simulations reach quasi-steady state for all resolutions, with visible seasonal trends in the region above 1 mbars (Fig.~\ref{fig:sidx_comp}). Simulations with higher resolution produces larger superrotation indices in the stratosphere, but they are all smaller than those in the TitanWRF simulations with flat surface, {\it e.g.}, 11 in $1^\circ$ simulation vs 16 in $5^\circ$ TitanWRF simulations. 

The AAM change rate $\partial{M}/\partial{t}$, which describes how rapidly angular momentum exchanges between the equatorial region ($\pm 15^\circ$) and the mid-to-high latitudes, also behaves differently when changing model resolutions. Figure~\ref{fig:am_comp} shows the AAM change rates in the stratosphere. In all cases, the AAM change rates show noticeable short-term variations relative to their seasonal trends; these variations are more prominent in the $2^\circ$ case, exhibiting strong oscillations with sharp peaks, especially in the equatorial region. These spikes with comparable magnitudes in both the equatorial region and the mid-to-high latitudes resembles the characteristics of "angular momentum transfer events" in TitanWRF \citep{Newman2011}. The $1^\circ$ and $4^\circ$ cases, on the other hand, show more gradual, year-around AM transfers between the hemispheres, which are similar to those in TAM GCM \citep{Lora2015,Lewis2023TAMWaves}. 

\subsection{Seasonal variations: temperature, wind and methane}

\subsubsection{Temperature and wind}

\begin{figure}
\centering
\includegraphics[width=0.8\linewidth]{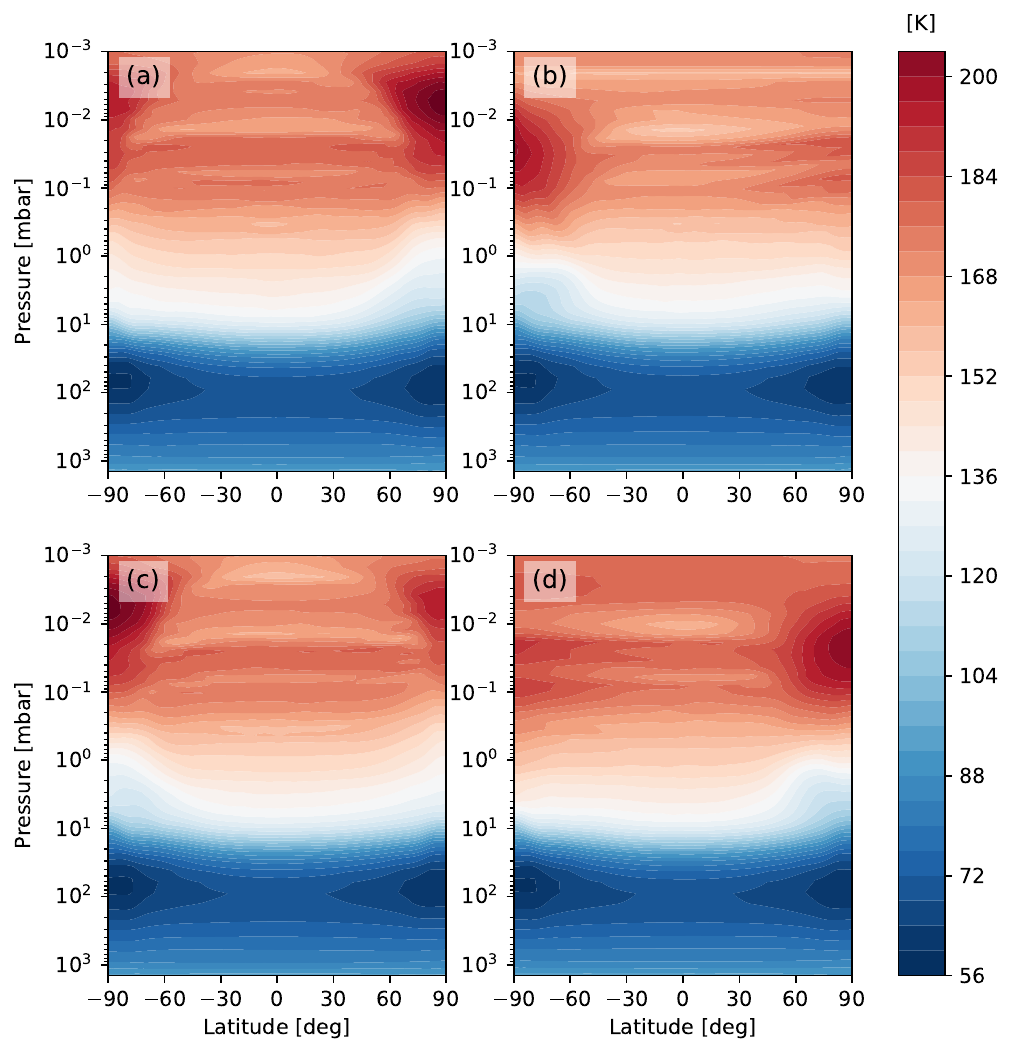}
\caption{\label{fig:t_zm}Zonal mean temperature near spring equinox $L_s=0^\circ$ (a), northern summer $L_s=90^\circ$ (b), autumn equinox $L_s=180^\circ$ (c) and northern winter $L_s=270^\circ$ (d).}
\end{figure}

The temperature and wind simulated by TitanMPAS exhibit seasonal patterns similar to those in previous GCM studies. The former can be further compared to the CIRS observations directly, while the latter may be compared to the wind derived from the CIRS temperature field \citep{Flasar2005} and the ALMA observations \citep{Cordiner2026}. Figure~\ref{fig:t_zm} shows the zonal mean temperature during the four seasons $L_s=0^\circ,\, 90^\circ,\,180^\circ,\, {\rm and}\, 270^\circ$. During both equinoxes (Fig.~\ref{fig:t_zm}a and Fig.~\ref{fig:t_zm}c), the stratosphere experiences warming at both poles above 0.1 mbar, with the north pole being slightly warmer than the south pole during $L_s=0^\circ$ and vice versa for $L_s=180^\circ$.  During northern summer and winter (Fig.~\ref{fig:t_zm}b and Fig.~\ref{fig:t_zm}d), the south pole and north pole become noticeably warmer than their respective counterparts.  The magnitude of the polar warming ($\rm \sim 205-210\,K$) is comparable to those in the CIRS observations \citep{Achterberg2023,Mathe2020} and TitanWRF predictions \citep{Newman2011}. The notable difference between TitanMPAS and TitanWRF predictions lies in the tropospheric temperature. TitanMPAS produces $\rm\sim 70\, K$ tropopause temperature at $\rm 100 \, mbar$ in the equatorial region with a mean near-surface temperature of about $\rm 93\, K$, whereas TitanWRF produced a cooler tropopause ($\rm \sim 67\, K$) at a lower altitude ($\rm \sim 200\, mbar$), as well as a cooler near-surface temperature of about $\rm  90\,K$ \citep{Newman2011,Newman2016}. The poles are distinctively colder than the equatorial region near the tropopause, a feature that presents in both TitanMPAS and TitanWRF year around but does not seem to be apparent in TAM and IPSL/PCM \citep{Lora2015, Lebonnois2012a,DeBatzDeTrenquelleon2025}. Note that, while the warm layer near 0.1 mbar results from the radiative heating of detached haze layer, the oscillatory temperature pattern above 0.1 mbar is caused by the wave reflection from the upper solid boundary in TitanMPAS, where the Rayleigh damping is insufficient to suppress all waves. Models with a free upper boundary like TitanWRF usually do not suffer from the same issue.

\begin{figure}
\centering
\includegraphics[width=0.8\linewidth]{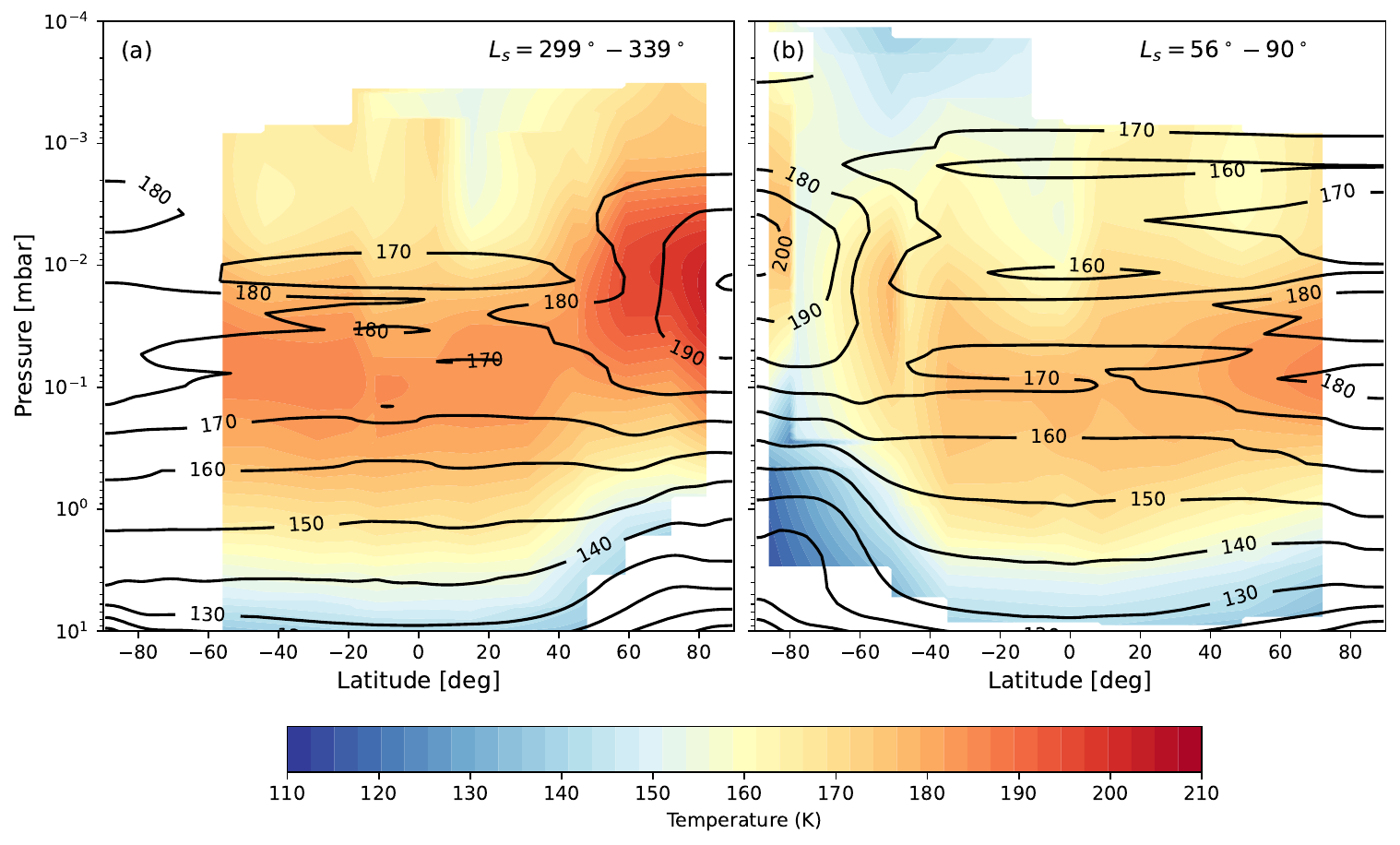}
\caption{\label{fig:t_zm_cirs}Zonal mean temperature during $L_s=299^\circ-339^\circ$ (a) and $L_s=56^\circ-90^\circ$ (b). The shaded color maps are the CIRS observations, reproduced using the data from \citet{Mathe2020}. The black contour lines are TitanMPAS model predictions.}
\end{figure}

The overall zonal mean temperature predicted by TitanMPAS matches well with the CIRS observations in the stratosphere. Figure~\ref{fig:t_zm_cirs} shows the comparison between the observed  (reproduced from \citet{Mathe2020}) and the predicted temperature. The pressure levels and the magnitude of polar warming and cooling including their meridional extension match closely. The temperature at low-to-mid latitudes in the region deeper than 0.1 mbar is also comparable between the model predictions and the observations. The notable mismatch occurs in the region above $\rm 10^{-3}\, mbar$, where the CIRS observations show about 10 K cooler temperature than the model predictions. However, this pressure level is approaching the sponge layer, where the Rayleigh damping has increasing impact on wind and temperature towards the model top. Compared to those in Fig.~\ref{fig:t_zm}, the long-time averaged temperature shows less oscillations in the region above 0.1 mbar at similar latitudes, implying that these oscillations are transient.

\begin{figure}
\centering
\includegraphics[width=0.8\linewidth]{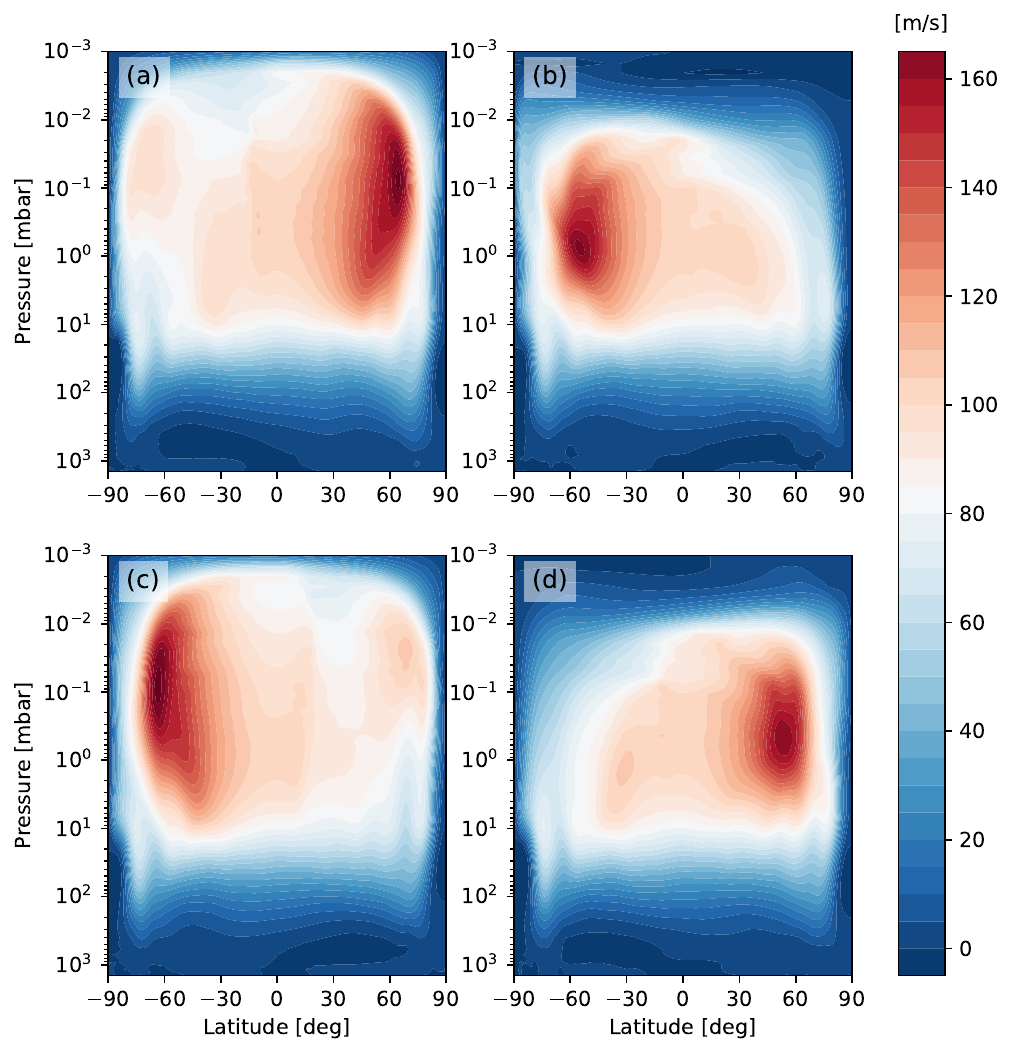}
\caption{\label{fig:u_zm}Zonal mean zonal wind near spring equinox $L_s=0^\circ$ (a), northern summer $L_s=90^\circ$ (b), autumn equinox $L_s=180^\circ$ (c) and northern winter $L_s=270^\circ$ (d).}
\end{figure}

The zonal mean zonal wind in TitanMPAS exhibits seasonal patterns similar to other GCM predictions but with noticeable differences in terms of how the peak wind speed distributes along latitude and pressure. Figure~\ref{fig:u_zm} shows that the zonal mean zonal wind in the stratosphere peaks at both northern and southern high latitudes during the equinoxes, with the northern maximum far exceeding its southern counterpart in northern spring ($L_s=0^\circ$), and vice versa in northern autumn ($L_s=180^\circ$). During northern summer and winter ($L_s=90^\circ$ and $L_s=270^\circ$), the zonal mean zonal wind peaks only at the southern and northern mid-to-high latitudes, respectively. The overall zonal wind magnitude is slightly weaker than those predicted in the TitanWRF with flat surface, {\it e.g.}, $160-170\,\rm ms^{-1}$ in TitanMPAS vs $180-200\,\rm ms^{-1}$ in TitanWRF depending on the seasons, but it is generally faster than those in TAM and IPSL/PCM. The vertical extent of the stratospheric superrotation stretches to a larger pressure range (1000 mbar  to 0.001 mbar) compare to that in TitanWRF, with a weaker subrotation near the model top ({\it e.g.}, retrograde wind with magnitudes $<10\,\rm ms^{-1}$ in TitanMPAS vs $20-30\,\rm ms^{-1}$ in TitanWRF). The equatorial superrotation presents throughout the year, but its magnitude relative to the peak wind speed is smaller than that in TitanWRF and IPSL/PCM \citep{Newman2011, DeBatzDeTrenquelleon2025}. So far, among all popular Titan GCMs, only TitanWRF with flat surface is able to predict the CIRS retrieval of peak wind speed near $30^\circ$ N during southern summer.

Despite the similarity among the GCM predicted zonal wind and those derived from the CIRS temperature measurements, TitanMPAS and other GCMs do not capture the zonal wind peaks observed by ALMA in the equatorial region \citep{Cordiner2026}. The equatorial zonal wind cannot be reliably retrieved from the CIRS temperature measurements since the gradient wind equation breaks down at the low latitudes \citep{Flasar2005}. However, a recent study on the stratospheric wind between 150 km and 450 km ($\sim \rm 100\, Pa - 0.1\, Pa$) by ALMA observations shows that the peak zonal wind locates near $10^\circ$ S during early northern summer and shifts towards the equator in the late northern summer \citep{Cordiner2026}. Neither TitanMPAS nor other Titan GCMs are able to correctly produce the peaks of the zonal wind during this period, suggesting insufficient equatorial acceleration by eddies relative to the deceleration by the advection of angular momentum in the same region \citep{Lombardo2023, Lewis2023TAMWaves, Lian2025}.
 
\subsubsection{Seasonal methane cycle: methane evaporation and precipitation}

\begin{figure}
\centering
\includegraphics[width=0.8\linewidth]{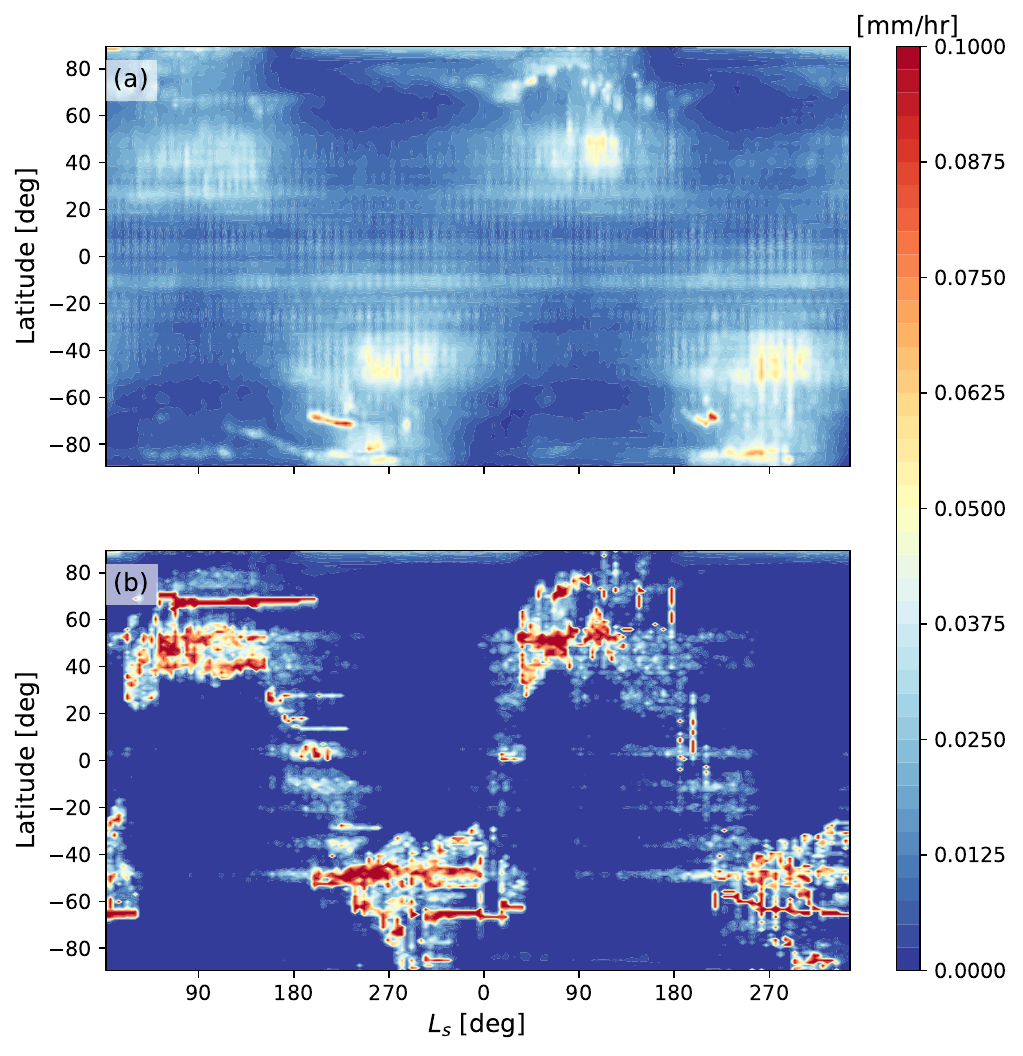}
\caption{\label{fig:qfx_cuprec}Zonal mean surface methane flux rate (due to evaporation) and precipitation rate over two consecutive Titan years.}
\end{figure}

The methane cycle shows seasonal patterns similar to those in the previous GCM predictions and the observations, but details differ. Figure~\ref{fig:qfx_cuprec} shows the seasonal variation of surface methane evaporation rate and precipitation rate as a function of latitude and solar longitude, the rates are in unit of millimeter per Earth hour. The surface evaporation rate is proportional to the near surface wind and saturation vapor pressure, it occurs at nearly all latitudes during most of the time in a Titan year depending on the liquid surface methane availability (Fig.~\ref{fig:qfx_cuprec}a). There is an apparent north-south asymmetry in the surface evaporative fluxes, which show relatively uniform year-round evaporation in the northern hemisphere and more intensified  evaporation in the southern mid-to-high latitudes during the southern summer. This occurs despite the fact that the prescribed permanent large lakes are located mostly in the northern high latitudes. Such behavior is consistent with the longer northern summer but shorter and warmer southern summer due to Titan's obliquity, as shown in previous studies \citep{Schneider2012,Lora2015,Lora2015b,Newman2016,Tokano2019}

The precipitation rate exhibits similar north-south asymmetry in terms of the intensity (Fig.~\ref{fig:qfx_cuprec}b). Unlike surface evaporation, the overall precipitation rate appears to be more sporadic both spatially and temporally. The precipitation in the northern high latitudes ({\it e.g.,} poleward of $60^\circ$ N)  initiates in the mid northern spring and lasts until northern autumn. The precipitation in the southern high latitudes, on the other hand, has a shorter duration. It initiates in the southern summer and last until mid southern autumn. The durations of precipitation in the mid latitudes are more comparable between the two hemispheres, with both last about $\Delta L_s=150^\circ$. At low latitudes, the precipitation patterns follow the locations of inter-tropical convergence zone (ITCZ) that passes from the summer hemispheres across the equator near equinoxes ({\it e.g.,} \citet{Newman2016}).

The detailed precipitation patterns predicted by TitanMPAS may be compared to other GCM predictions and the observations directly, though each GCM has different configuration in terms of surface methane and physics parameterization of methane cycle. The precipitation patterns predicted by TitanMPAS stand as a middle ground among the cases with the infinite global surface methane reservoir and the finite surface methane availability \citep{Newman2016, Lora2015, Lora2019}. In the cases with global methane reservoir, the precipitation occurs year-around with continuous intertropical convergence zone (ITCZ) from summer hemisphere to winter hemisphere \citep{Newman2016,Lora2015}. Whereas in the cases with finite surface methane reservoir, the precipitation is limited to high latitudes in the summer hemispheres \citep{Newman2016}. TitanMPAS predicts precipitation in a wider region that covers the mid-to-high latitudes in the summer hemispheres, as well as trace amount of precipitation at low latitudes during the equinoxes. The overall distribution of precipitation is more similar to the TAM case with the subsurface methane model \citep{Lora2022} except with a few times larger precipitation rates. 

The precipitation patterns predicted by TitanMPAS may be used to interpret the observed cloud coverage from southern summer to northern summer. We clarify that the precipitation only indicates locally generated clouds because the moist convection scheme in TitanMPAS (likewise in TAM) does not store cloud particles in the atmosphere. In reality, clouds are subjected to advection by wind, meaning that cloud cover is not necessarily co-located with precipitation. Moreover, the predicted precipitation does not represent other types of cloud formation such as stratiform clouds or clouds associated with wave activities.

The seasonal and spatial distribution of precipitation are similar to the observations of cloud activities in the southern hemisphere during the southern summer but mismatch with those in the northern hemisphere during early northern spring. The precipitation occurs at the south pole near the southern summer solstice and extends to southern mid-to-low latitudes as the season progresses to the northern spring equinox. The pattern is consistent with the ground-based telescope and Cassini observations during the same time period \citep{Bouchez2005,Rodriguez2011,Schaller2006,Turtle2011}. A major mismatch between the predicted precipitation and the observed cloud activities occurs during late southern summer and early northern spring where the observations show cloud coverage in the $60^\circ$ N polar cap \citep{Turtle2011}, whereas TitanMPAS predicts very little precipitation. The precipitation coverage at mid-latitudes after the northern spring equinox is similar to the observed cloud activities \citep{Lora2015}. Frequent cloud activities (including large, multi-week convective storms) were observed at northern mid-to-high latitudes during the northern summer by ground-based and space telescopes \citep{Corlies2022DPS,Nixon2025}, coinciding with the intensified precipitation predicted by TitanMPAS during the same time period.

\subsubsection{Atmospheric methane transport}
\label{methane_transport}

\begin{figure}
\centering
\includegraphics[width=0.8\linewidth]{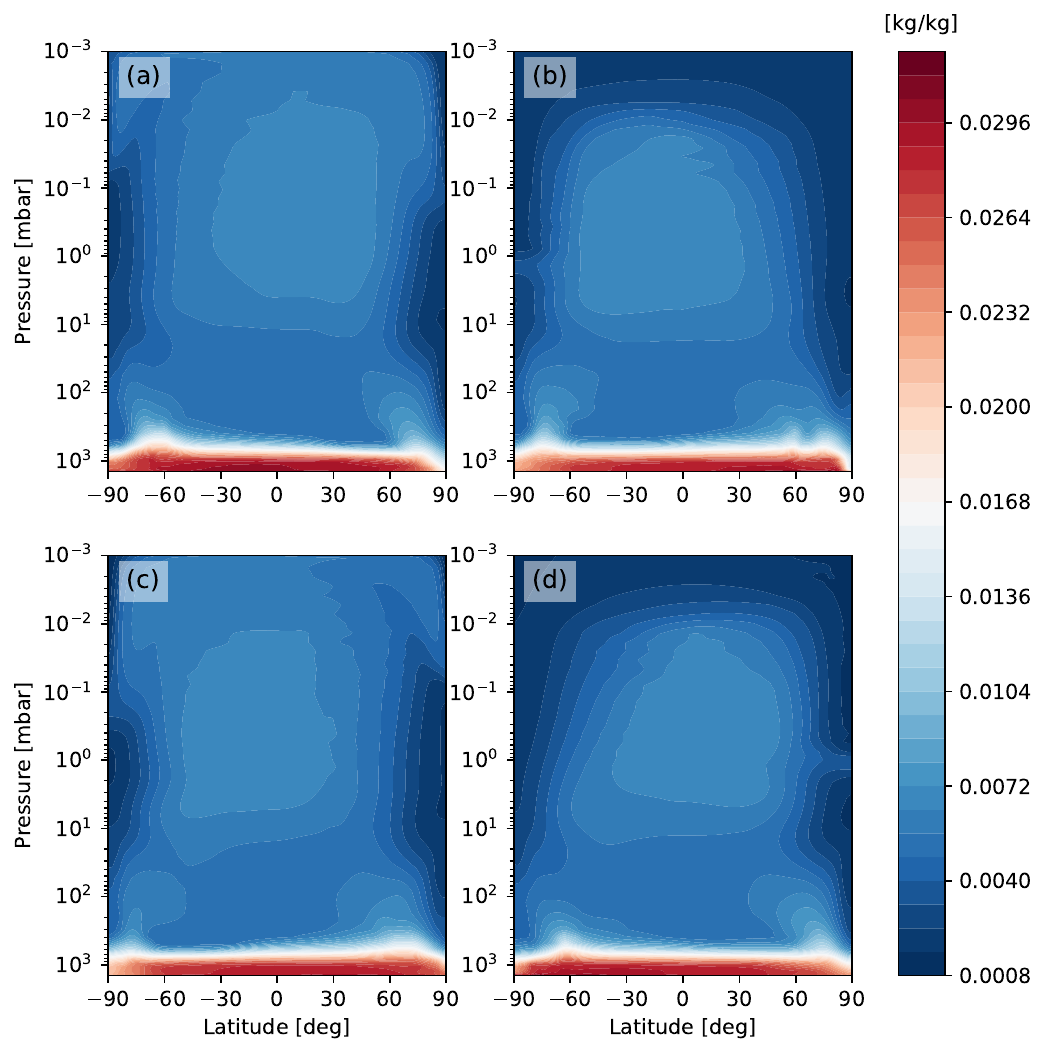}
\caption{\label{fig:qv_zm}Zonal mean methane vapor mass mixing ratio near spring equinox $L_s=0^\circ$ (a), northern summer $L_s=90^\circ$ (b), autumn equinox $L_s=180^\circ$ (c) and northern winter $L_s=270^\circ$ (d).}
\end{figure}

The zonal mean methane vapor mass mixing ratio does not exhibit strong seasonal variation as the surface evaporation and precipitation. Here the mass mixing ratio $q$ is the direct model output and it is related to the volume mixing ratio $\chi$ as $q=\chi\frac{m_{N2}}{m_{air}}$, where $m_{CH_4}=16$ is the molecular mass of methane and $m_{air}=28.6$ is the mean molecular mass of air.  Figure~\ref{fig:qv_zm}  shows the distribution of methane vapor during the equinoxes and solstices. Methane vapor abundance is concentrated within a few hundred millibars above the surface, and decreases rapidly towards the tropopause where there is a layer of methane with local minima (between $\rm 10-500\, mbars$). In the region with  $\rm p< 10\, mbars$, methane vapor has relatively higher abundance at low-to-mid latitudes than that at high latitudes in both hemispheres. The inversion of methane abundance from the troposphere to the stratosphere $-$ characterized by a methane minimum between the upper troposphere and the lower stratosphere that transitions into an enhanced abundance above $\rm 10 \, mbar$ $-$ is consistent with the reanalysis of Cassini Visual and Infrared Mapping Spectrometer (VIMS) data between 80 km and 200 km by \citet{Rannou2021}. Additionally, the latitudinal variation of methane abundance simulated in our model is similar to that in the VIMS data, which showed methane was generally more abundant in the equatorial region than that in the high latitudes above 80 km.  The overall small variations of methane vapor distribution throughout the Titan year imply that the methane vapor may have slow exchanges between the troposphere and the stratosphere.

\begin{figure}
\centering
\includegraphics[width=0.8\linewidth]{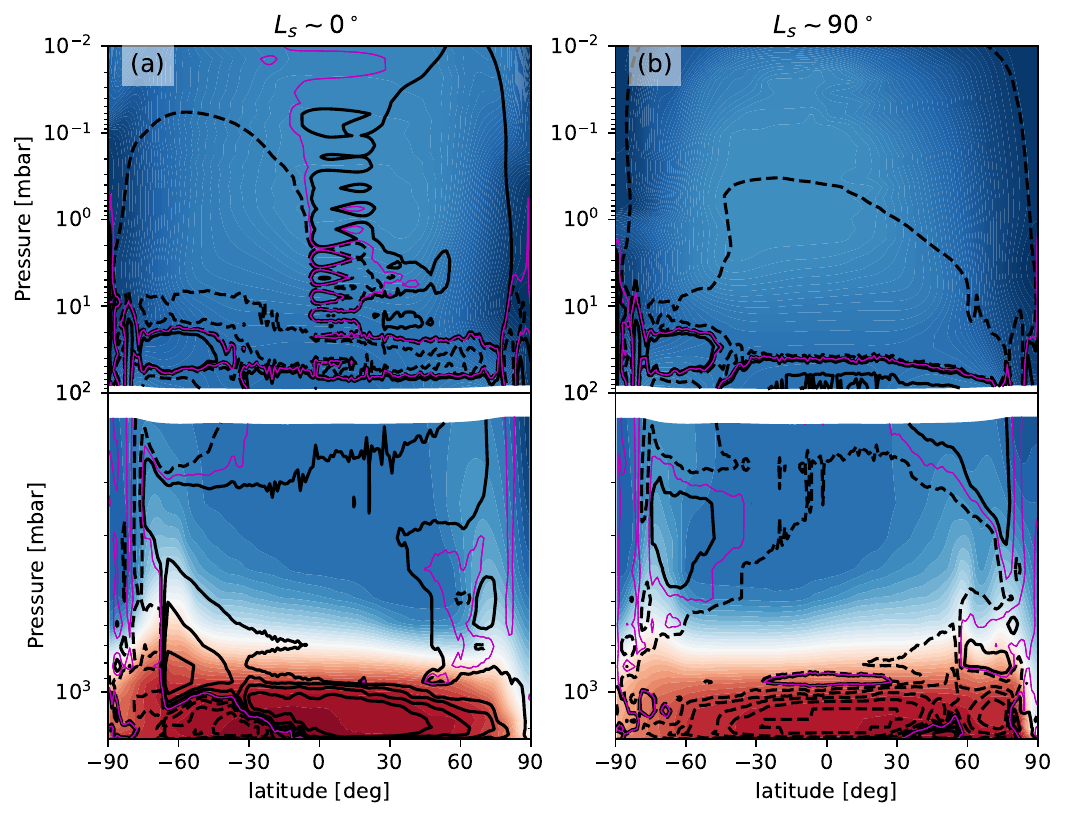}
\caption{\label{fig:psis_zm}Residual mean circulation $\overline\psi^*$ (black contour lines) and zonal mean methane vapor mass mixing ratio (shaded color maps, same as those in Fig.~\ref{fig:qv_zm}a and Fig.~\ref{fig:qv_zm}b respectively) near spring equinox $L_s=0^\circ$ (a) and northern summer $L_s=90^\circ$ (b). $\overline\psi^*$ is time-averaged over two Titan days. The troposphere and stratosphere are separated to better demonstrate the streamfunctions. The solid black contour lines mean clockwise rotation and the dashed black contour lines mean counterclockwise rotation. The solid magenta lines are the contours of $\overline\psi^*=0$.  The contour lines are separated by exponential spacing, which decreases by a factor of 2 in the troposphere and factor of 10 in the stratosphere. The maximum residual mean circulation is on order of $\rm 10^3 \,kgm^{-1}s^{-1}$. The primary circulation patterns during autumn equinox $L_s=180^\circ$ and southern summer $L_s=270^\circ$ (not shown) are similar to those of $L_s=0^\circ$ and $L_s=90^\circ$ respectively, except that the rotation directions are reversed.}
\end{figure}

The spatial and temporal distribution of methane vapor abundance is a result of combined effect from methane cycle, dynamic transport (mean and eddy) and ad-hoc damping in the stratosphere with damping time scale decreases from 5 Titan years near the tropopause to 100 Titan days near the model top. The long damping time scale, especially in the lower stratosphere, is to minimize the impact of ad-hoc relaxation on the dynamic mixing, since the intention of the damping is to suppress the pile-up of methane vapor near the model top where the Rayleigh damping weakens the lateral and vertical mixing. Figure~\ref{fig:psis_zm} shows the advective methane transport represented by residual mean circulation $\overline\psi^*$ near spring equinox ($L_s\sim 0^\circ$) and northern summer ($L_s\sim 90^\circ$) as examples. The residual mean circulation shows stacked circulation cells below and above the pressure level $\rm p\sim 50\,mbar$, which are separated by the near-zero pole-to-pole streamfunctions with flat appearance (solid magenta lines in Fig.~\ref{fig:psis_zm}) . During both seasons, methane with high abundance is trapped in the lower troposphere (roughly $\rm p> 1000\,mbar$)  and well mixed by strong meridional circulations, {\it i.e.}, a pair of circulation cells arising near $15^\circ$ S during $L_s\sim 0^\circ$ and a single pole-to-pole cell during $L_s\sim 90^\circ$.  The lower methane abundance between $\rm 50-500\,mbars$ is mostly a result of downward mixing of methane by advection.  Above $\rm 50 \, mbars$, the methane transport is dominated by a pair of  nearly zonally symmetric circulation cells during $L_s\sim 0^\circ$ and again a single pole-to-pole counterclockwise circulation cell during $L_s\sim 90^\circ$. While the methane transport at the low-to-mid latitudes is mostly isolated to its respective tropospheric and stratospheric regions, the secondary polar cells poleward of $70^\circ$ in both hemispheres can transport methane from the lower atmosphere (above a few hundred millibars) to about $\rm 10\,mbars$. Subsequently, the lateral transport driven by the pole-to-equator residual circulation branch may be responsible for the enhanced methane abundance at low-to-mid latitudes above $\rm 10 \, mbars$.

\begin{figure}
\centering
\includegraphics[width=0.8\linewidth]{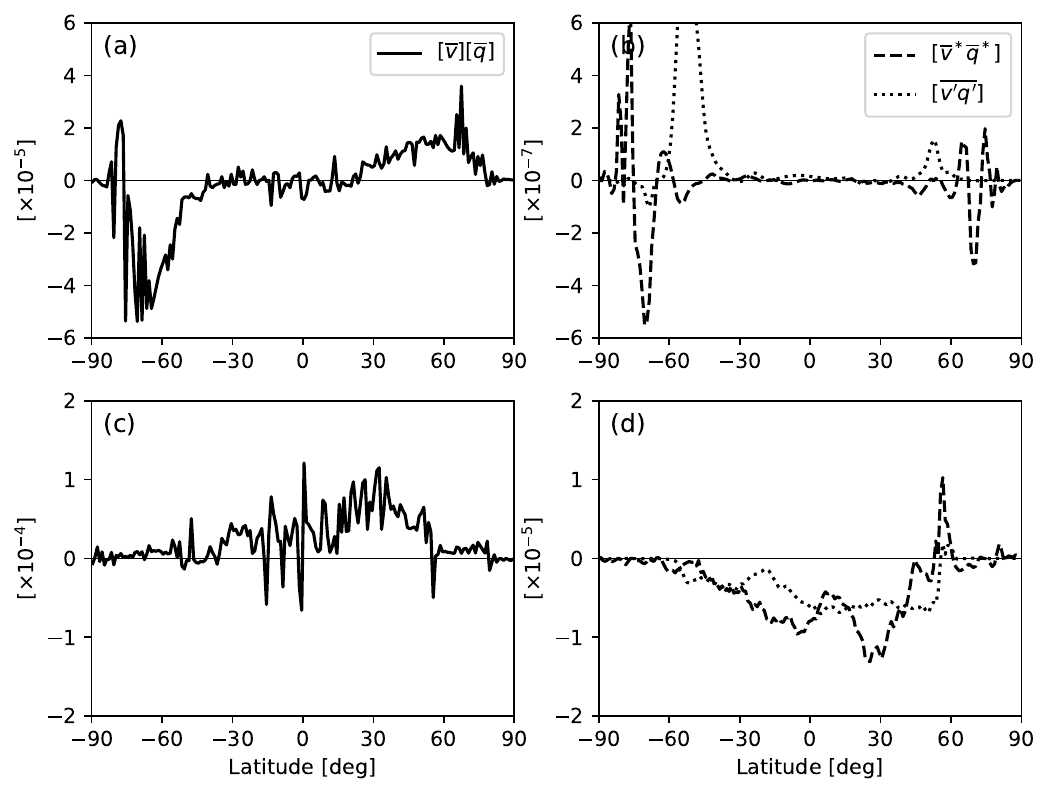}
\caption{\label{fig:vq_covariance}Meridional transport of methane vapor due to mean advection $[{\overline{v}}][{\overline{q}}]$ (a, c), stationary eddies $[{\overline{v}}^*{\overline{q}}^*]$ and transient eddies $[{\overline{v^\prime q^\prime}}]$ near $L_s=90^\circ$ (b, d). (a, b) and (c,d) are mass-weighted vapor fluxes in the stratosphere and troposphere (above and below $\rm p\sim 100\,mbars$), respectively. The unit for these quantities is $[ms^{-1}\cdot kg/kg]$.}
\end{figure}

The residual mean circulation is close to the Eulerian mean circulation (not shown) in the majority of the atmosphere, suggesting weak eddy contribution to the methane transport. Similar findings also exist in TAM model results \citep{Lewis2023TAMWaves}. We decompose the mean and eddy contribution to methane transport using the method describe in \citet{Peixoto1992} to demonstrate the relative importance between mean and eddy transport. The decomposition splits the total tracer transport to three components including mean advection, stationary eddies and transient eddies :
\begin{equation}
    \overline{[AB]} = [{\overline{A}}][{\overline{B}}] +[{\overline{A}}^*{\overline{B}}^*]+[{\overline{A^\prime B^\prime}}]
\end{equation}
where $A$ and $B$ are physics quantities such as ($v, q$) or ($w,q$) with $v$ and $w$ being meridional and vertical velocity and $q$ being methane mixing ratio. $A^\prime=A-{\overline{A}}$ is the perturbation with respect to the time average $\overline{A}$ and $A^*=A-[A]$ is the perturbation with respect to the zonal mean $[A]$. Figure~\ref{fig:vq_covariance} shows the meridional transport of methane vapor at selected pressure levels with their transport due to mean advection $[{\overline{v}}][{\overline{q}}]$, stationary eddies $[{\overline{v}}^*{\overline{q}}^*]$ and transient eddies $[{\overline{v^\prime q^\prime}}]$ near $L_s=90^\circ$. The meridional convergence and divergence of these components correspond to the increase or decrease trend of methane abundance respectively.  All meridional methane fluxes are mass-weighted and averaged in either the troposphere or the stratosphere, {\it e.g., } $\int_{p_b}^{p_t}[{\overline{AB}}]dp/\int_{p_b}^{p_t}dp$, where $p_b$ and $p_t$ mean the bottom and the top of the vertical domains.

The mean advection (Fig.~\ref{fig:vq_covariance}a, c) dominates the methane transport as it is $1-2$ orders of magnitude larger than the eddy fluxes (Fig.~\ref{fig:vq_covariance}b, d) in both the stratosphere and the troposphere. In the stratosphere ($\rm p<100\,mbars$), the mean advection $[{\overline{v}}][{\overline{q}}]$ (Fig.~\ref{fig:vq_covariance}a) shows poleward fluxes of methane in both hemispheres, and it is about 2 orders of magnitude larger than the eddy fluxes (Fig.~\ref{fig:vq_covariance}b).  The transient eddies $[{\overline{v^\prime q^\prime}}]$ dominates the eddy transport at the mid-latitudes, while the stationary eddies  $[{\overline{v}}^*{\overline{q}}^*]$ dominates the high latitudes. The divergence of $[{\overline{v}}][{\overline{q}}]$ implies a loss of methane at low-to-mid latitudes. However, the amount of net meridional flux depends on how we choose $p_b$ and $p_t$, as the mean circulation patterns between $\rm 10-500\, mbars$ are complicated (Fig.~\ref{fig:psis_zm}). In the troposphere,  $[{\overline{v}}][{\overline{q}}]$ shows an net south-to-north flux, implying that it is dominated by the returning branch of the cross-equator circulation cell near the surface (lower panel of Fig.~\ref{fig:psis_zm}b). This net flux may be countered by surface methane fluxes and moist convection, as well as other methane transport such as the PBL mixing. The stationary eddy fluxes are about an order of magnitude smaller than the mean advection, but comparable or larger than the transient eddy fluxes at nearly all latitudes. The comparison among all meridional methane fluxes in the stratosphere and the troposphere suggests that the topography, a major source of stationary eddies, has a strong influence on the eddy transport in the troposphere as well as the high latitudes in the stratosphere. The relative importance between  $[{\overline{v}}^*{\overline{q}}^*]$ and $[{\overline{v^\prime q^\prime}}]$ are seasonal dependent (not shown), but their strengths are comparable throughout the Titan year. 

\citet{Lora2015b} showed that transient eddies in the troposphere might be critical in generating the north-south asymmetry of liquid surface methane. However, their model did not include the topography at the time therefore the influence of the stationary eddies were negligible. \citet{Lora2022} studied methane transport with topography but implemented subsurface methane model as opposed to the prescribed polar liquid methane reservoirs in \citet{Lora2015b}. They determined that the north-south asymmetry of lake distribution was unlikely caused by topography though did not rule out other possibilities such as orbital forcing and surface properties related to surface methane availability. These studies pose new challenges in understanding the roles of topography and stationary eddies on producing the north-south asymmetry of methane lake distribution, as the contributions from transient and stationary eddies can be comparable in our model results, perhaps due to the higher grid resolution that tends to amplify stationary eddies {\it e.g.}, \citet{Yamamoto2021,Niekerk2020}.

\subsubsection{Comparison to Huygens measurements}

\begin{figure}
\centering
\includegraphics[width=1\linewidth]{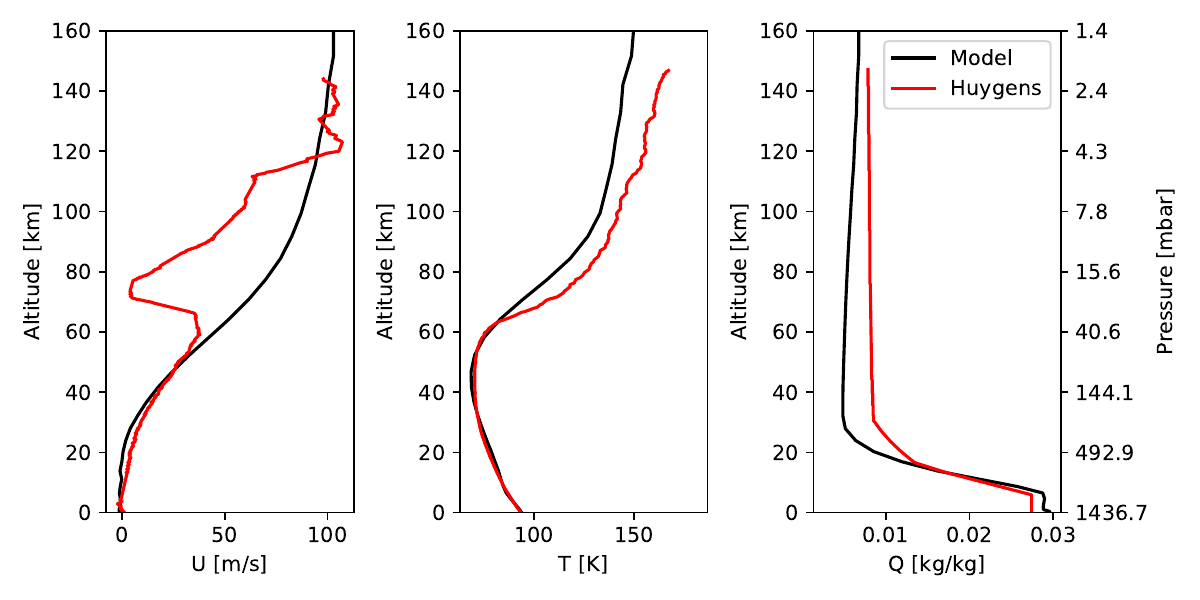}
\caption{\label{fig:1D_Huygens}Zonal wind, temperature and methane mass mixing ratio as a function of altitude/pressure near ($167^\circ$ E, $10^\circ$ S) and $L_s\sim303^\circ$, when Huygens probe performed in-situ measurements.}
\end{figure}

The wind, temperature and methane abundance predicted by TitanMPAS compare reasonably well against the only in-situ measurements by Huygens probe to date \citep{Bird2005,Fulchignoni2005,Niemann2010}. Figure~\ref{fig:1D_Huygens} shows the comparison between the modeled and measured profiles near Huygens landing site ($167^\circ$ E, $10^\circ$ S) and $L_s\sim 303^\circ$. The modeled and the observed zonal wind are very close in the troposphere, with the former being slightly slower ($\rm <1ms^{-1}$) than the latter between 20 km and 40 km. The wind profiles also show good match above 120 km, where the wind speed is close to be $\rm 100\, ms^{-1}$. Like most other Titan GCMs, TitanMPAS does not capture the observed wind minimum near 75 km. Note that the IPSL/PCM GCM relies on nudging the tropospheric wind towards the Huygens wind profile, therefore imposed the wind minimum near 75 km \citep{DeBatzDeTrenquelleon2025}.  The modeled temperature is also in good agreement with the measurement in the troposphere, but becomes relatively cooler in the stratosphere (above 60 km), whereas other Titan GCMs tend to produce slightly warmer stratosphere. In the lower 20 km, the modeled methane mass mixing ratio matches well with the measurement, capturing the near-constant value within 10 km above the surface. However, the model underestimates the measurements slightly between 20 km and 140 km. Overall the tropospheric wind, temperature and methane profiles predicted by TitanMPAS are on-par or better than those predicted by other popular Titan GCMs ({\it e.g.}, \citet{Lora2019}).

\subsection{Dynamic properties near Huygens site and Selk crater}

The good match between our model results and the Huygens probe measurement suggests that TitanMPAS is suitable to simulate the dynamics in the lower troposphere. This capability may benefit the future Dragonfly mission \citep{Lorenz2018,Lorenz2021}, which will provide in-situ measurement of wind, temperature and tracer abundances near Selk crater that is centered at $160^\circ$ E and $7^\circ$ N. Here we present preliminary predictions of near-surface dynamic properties on diurnal timescale. The detailed dynamics at the location will be further investigated in our regional mesoscale simulations with typical 1 km horizontal resolution, which well resolves the Selk crater features including the rims that are only a few tens of kilometers wide.

\begin{figure}
\centering
\includegraphics[width=1\linewidth]{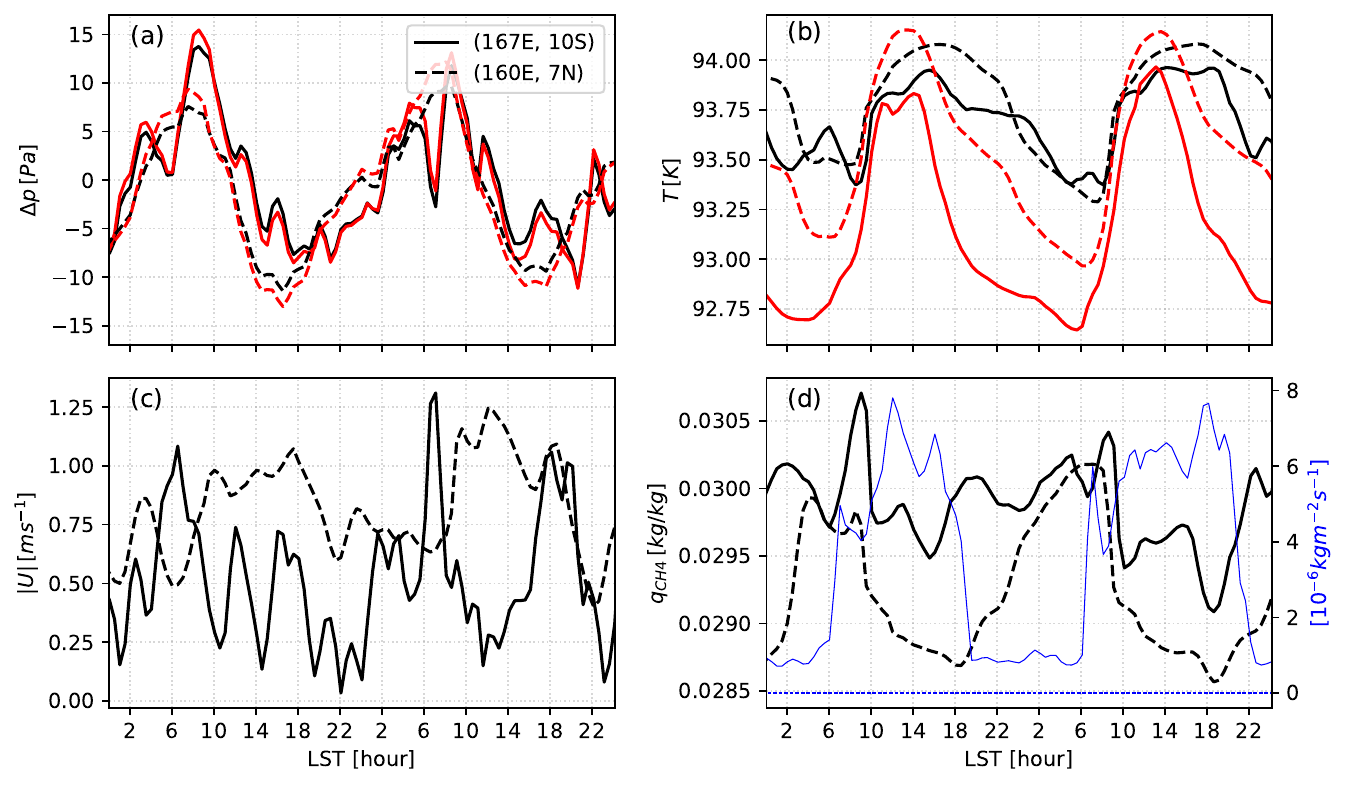}
\caption{\label{fig:daily_vars}Air/surface pressure anomaly (a), air/surface temperature, wind magnitude and methane vapor abundance near $167^\circ$ E, $10^\circ$ S (solid lines) and   $160^\circ$ E, $7^\circ$ N (dashed lines) as a function of local time over two consecutive days.  The season is at $L_s\sim303^\circ$, when Huygens probe performed in-situ measurements. The black lines are for dynamic quantities in the air, and the red lines are for the quantities at the surface. The thin blue lines in panel (d) are the surface methane evaporate flux.}
\end{figure}

Figure~\ref{fig:daily_vars} shows the pressure anomaly, temperature, wind magnitude and methane vapor mass mixing ratio near both Huygens landing location ($167.7^\circ$ E, $10^\circ$ S) and Selk crater ($160^\circ$ E, $7^\circ$ S) during $L_s\sim 303^\circ$.  All dynamic quantities shown are in the lowest model layer (about 150 m above the surface) except those for the surface (red lines). The anomalies of  air pressure and the surface pressure show tight correlations, and they vary about $10-15$ Pa diurnally. The small diurnal variation of pressure is expected given the weak planet rotation and wind in the lower troposphere. We clarify that the simulation results presented here are from the model without the parameterized tidal forces from Saturn \citep{Charnay2022}, although such parameterization has now been implemented in a newer version of TitanMPAS and it is still undergone tests for its long-term effect on global dynamics. Therefore, the pressure variations seen here are predominantly thermal tides with some topographic effects.

The surface temperature and air temperature behave quite differently. First, the peak air temperature lags behind the peak surface temperature by about two Titan hours. Second, while the air temperature is comparable between the two locations, the surface temperature is much cooler than the air temperature near the Huygens location compared to the Selk crater location (albeit the maximum difference is less than 1 K). This is due to the larger surface cooling rates caused by more surface methane evaporation near the Huygens site (the solid blue line vs the dashed blue line in Fig.~\ref{fig:daily_vars}d). This is expected as the evaporative surface methane fluxes in the southern low latitudes is larger than those in the northern low latitudes during most of the time (Fig.~\ref{fig:qfx_cuprec}a). Accordingly, the methane vapor abundance is also higher in the southern low latitudes than that in the northern low latitudes (Fig.~\ref{fig:daily_vars}d).

The magnitude of the near surface wind is smaller than $\rm 1\,ms^{-1}$ most of the time during a Titan day at both locations (Fig.~\ref{fig:daily_vars}c). This is consistent with the Huygens measurement, as well as other GCM predictions ({\it e.g.} \citet{Folkner2006, Lora2019}). The wind peaks near the early morning (6 AM) at the Huygens site and near the early afternoon (~2 PM) at the Selk crater. The orientation of the wind also varies between the locations (not shown). At the Huygens site, the wind blows eastward most of the time but briefly turns to westward during the afternoon (2 PM to 6 PM). At the Selk crater, the northerly-easterly (from northeast to southwest) wind dominates throughout the day. It should be noted that the wind magnitude and orientation described here are from the $1^\circ$ ($\rm\sim 45\, km$) simulations, which cannot resolve mesoscale features such as the Selk crater that has a diameter of less than a hundred kilometers. The wind may undergo dramatic change due to the geometry of the crater as well as the radiative effect that drives the slope wind ({\it e.g.}, \citet{Lefevre2026}), especially near the steep cater walls. 

\section{Conclusion and discussion}
\label{conclusion_discussion}

We present a new Titan GCM $-$ TitanMPAS, a global and regional circulation model based on MPAS and TitanWRF GCMs. This model is complement to planetMPAS that has simulated atmospheric circulations on Mars and Jupiter \citep{Lian2023}. TitanMPAS takes advantage of MPAS' unique unstructured mesh that is globally uniform, thus avoiding the polar convergence problem that plagues the GCMs implementing traditional spherical-polar grids ({\it e.g.}, TitanWRF). Besides the better conservation properties than the finite difference method utilized by TitanWRF, the finite-volume dynamical core is also more efficient to handle high-spatial resolutions and suitable for complex geometry such as regional mesoscale simulations compared to spectral dynamical cores \citep{Virieux2011, Karniadakis2005spectral}. In additional to the new dynamical core, we also introduce a bulk moist convection scheme, variable surface features (topography, albedo, thermal inertia and lake distribution) as well as a newly developed K-distribution radiative transfer model. Other improvements include the corrections made to the gravity and curvature terms due to the large aspect ratio between Titan's atmospheric depth to planet radius compared to other planets. Case studies of model resolution are also performed, which show higher wind speed can be achieved with higher horizontal resolution. 

Simulation results over more than 100 Titan years show that the model has reached a quasi-steady state. TitanMPAS is able to reproduce several key features of the atmospheric circulation patterns. Notably, it is able to produce the stratospheric superrotation with zonal wind speed exceeding $\rm 170\,ms^{-1}$ in the $1^\circ$ resolution simulation. Both the seasonal wind and temperature patterns predicted by TitanMPAS are similar to those in the observations and other Titan GCM predictions though details vary. The wind, temperature and vapor profiles also show good match with the in-situ measurements by Huygens probe near $L_s\sim 303^\circ$, particularly in the troposphere. All these dynamic quantities are better represented than those in TitanWRF, which tends to produce a cooler tropopause at a lower altitude and overly strong zonal wind in the troposphere \citep{Newman2011}.

The seasonal and spatial precipitation patterns predicted by TitanMPAS match well with the observed cloud activities throughout the Titan year except the time period when the late southern summer transits to early northern spring. The precipitation is more intense in the summer hemispheres with visible ITCZ patterns between the seasonal transitions. The seasonal variations of zonal mean methane vapor mass mixing ratio are less noticeable, which stays well mixed in the lower atmosphere ($\rm \sim 10\,km$ above the surface) and decreases towards the tropopause and then increases again in the lower stratosphere. The zonal mean methane mass mixing ratio maintains a relatively depleted layer between $\rm 500\, mbar$ and $\rm 10\,mbar$ (corresponding to $\rm \sim 20\,km$ and $\rm \sim 90\,km$ respectively) and enriched layer between $\rm 3\, mbar$ and $\rm 0.01\,mbar$ (corresponding to $\rm \sim 120\,km$ and $\rm \sim 360\,km$ respectively). The vertical variations of methane abundance resembles those in the reanalysis of the Cassini VIMS data. Overall, the mean circulation dominates the eddy transport of methane in the entire atmosphere, with the former being typically an order of magnitude larger than the latter.

Finally, we demonstrate diurnal variations of physics quantities including pressure, temperature, wind and methane abundance near the Huygens site and the Selk crater. The variations of pressure and temperature are very small relative to their mean values due to the weak radiation and near surface wind, but the diurnal cycles are clearly presented. Similar to the Huygens in-situ measurements and other GCM predictions, the near-surface wind is generally smaller than $\rm 1\,ms^{-1}$, and it is dominated by westerlies but the wind may blow from east to west in the afternoon near the Huygens site. Our regional mesoscale resolving model results of the Selk crater, which will be presented in Part II of TitanMPAS, show that the wind can reach $\rm\sim 2ms^{-1}$ occasionally at about 1 m above the surface and exceed $\rm 3\,ms^{-1}$ at $\rm\sim 160\,m$ above the surface, much faster than the wind speed predicted by the $1^\circ$ simulations near the similar altitude. This suggests that the $1^\circ$ resolution may not be sufficient to represent the local dynamics in the vicinity of Selk crater properly.

Despite the good match among the TitanMPAS, the observations and other Titan GCM results, several key features remain unexplained. First, what produces the observed wind minimum near 75 km? This feature may be global in the Cassini radio occultation results, that is, it exists from pole to pole \citep{Flasar2005}. \citet{Lebonnois2012a} demonstrated that, for sufficiently large lateral temperature contrast, the wind minimum might be achievable via thermal wind balance. In both TitanWRF and TitanMPAS, we observed similar behavior at lower altitudes during the early stage of the simulations when the zonal winds are generally weaker ($\rm u_{max}\sim 50\,ms^{-1})$. The wind minimum at mid-to-high latitudes disappeared once the wind became faster. In TitanMPAS, there are wind minima near $\rm 65\, km$ ($\sim\rm 30\,mbar$) within the 5-deg polar caps (Fig.~\ref{fig:u_zm}), which are produced by the year-around polar temperature minima in the regions (Fig~\ref{fig:t_zm}). To date, none of the Titan GCMs has been able to produce the lateral temperature contrast large enough to support the sharp vertical gradient of the zonal wind in the Huygens measurements. A possible candidate for the wind minimum near the low latitude is the momentum deposition by breaking gravity waves, {\it e.g.}, orographic waves \citep{Flasar2016,Flasar2020}. However, such mechanism requires further investigation since the presence of topography in TitanMPAS does not appear to impact the zonal wind near 75 km, although it may contribute to the slowdown of the stratospheric superrotation.

Another question is what mechanisms produce the north-south asymmetry of lakes. The observations of cloud activities suggested that the southern hemisphere had more intensive storms but the northern hemisphere had prolonged precipitation, which led to more lake formation in the northern high latitudes. \citet{Lora2015b} suggested that transient eddies might play an important role in producing the asymmetry. However, as we demonstrated, the stationary eddies may play an equally important role in the methane transport, as higher horizontal resolution in TitanMPAS produces larger terrain variance and tends to amplify orographic waves. A more recent study by \citet{Lora2022} suggested that the asymmetry might not be caused by topography but did not rule out the contribution from the orbital radiative forcing and the distribution of surface methane availability.  However, their results were obtained from simulations using low-resolution topography ($5.625^\circ$ or 250 km grid size, similar to that used by TitanWRF),  which may change when resolution increases, {\it e.g.,} $1^\circ$ grid size in TitanMPAS. 

\begin{figure}
\centering
\includegraphics[width=0.8\linewidth]{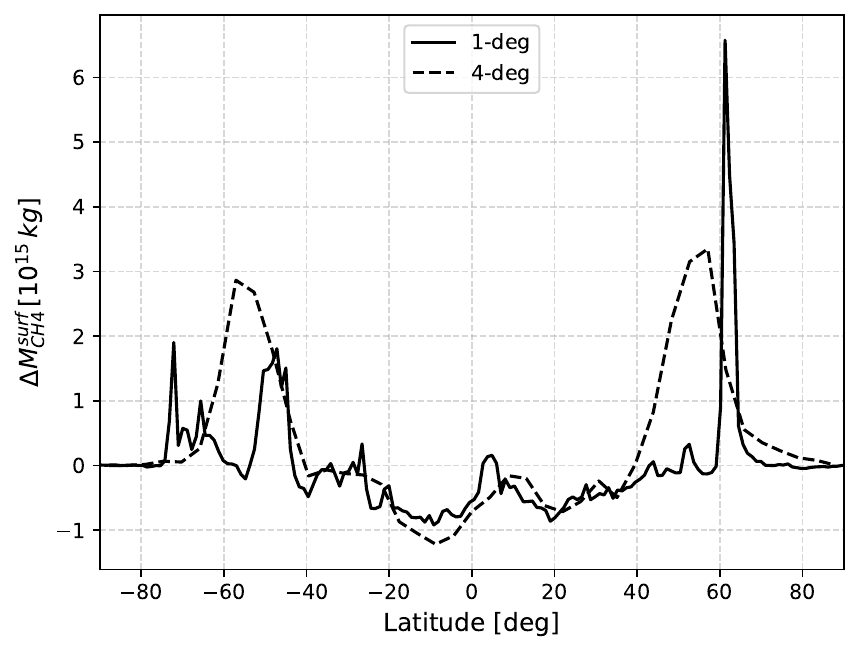}
\caption{\label{fig:surf_ch4} The change of liquid surface methane mass between the start ($L_s=0^\circ$) and the end ($L_s=360^\circ$) of a Titan year.  The solid line is the $1^\circ$ simulation result, and the dashed line is $4^\circ$ simulation result.}
\end{figure}

A quick comparison shows that the surface methane change due to the methane cycle is indeed affected by the model resolution.  Figure~\ref{fig:surf_ch4} shows the comparison of the liquid surface methane $\Delta M_{ch4}^{surf}$ between $1^\circ$ and $4^\circ$ simulations in TitanMPAS. $\Delta M_{ch4}^{surf}$ is the change of liquid methane mass between $L_s=0^\circ$ and $L_s=360^\circ$ of a Titan year (accumulated over a Titan year), which includes the contributions from both precipitation and surface evaporation. In $4^\circ$ simulations, $\Delta M_{ch4}^{surf}$ peaks near $\pm 50^\circ$ with comparable values but slightly larger in the northern hemisphere. In $1^\circ$ simulations, $\Delta M_{ch4}^{surf}$ shows a distinct peak in the northern high latitudes, which is more than twice as large as the peak value in the southern hemisphere. The latitudinal variation of $\Delta M_{ch4}^{surf}$ is also noticeably noisier in the $1^\circ$ simulations than that in the $4^\circ$ simulations. Despite the different distribution of $\Delta M_{ch4}^{surf}$ with different grid resolution, we cannot attribute the asymmetry to better resolved topography alone since higher grid resolution impacts overall dynamics, {\it e.g.}, Fig.~\ref{fig:sidx_comp}. The complexity of combined effect from methane cycle, topography, orbital forcing, the surface availability of liquid methane (such as lake distribution) and land hydrology ({\it e.g.}, \citet{Faulk2020}) warrants its own investigation.

\begin{figure}
\centering
\includegraphics[width=0.8\linewidth]{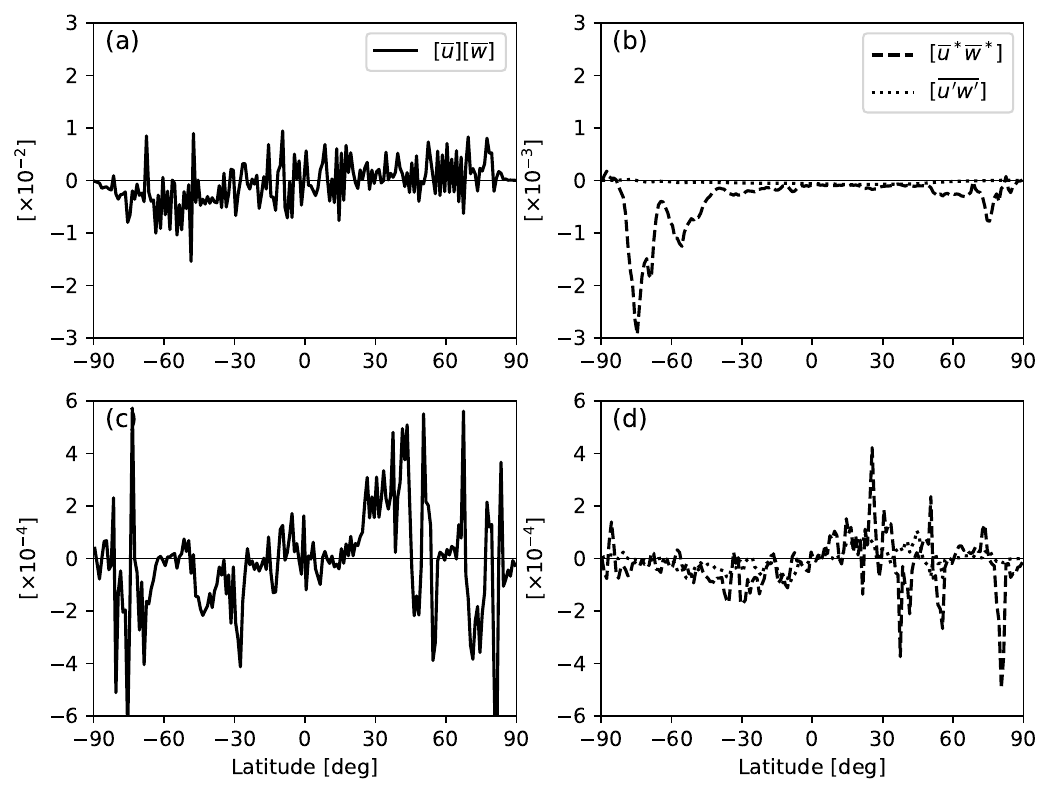}
\caption{\label{fig:uw_covariance}Vertical transport of zonal momentum due to mean advection $[{\overline{u}}][{\overline{w}}]$ (a, c), stationary eddies $[{\overline{u}}^*{\overline{w}}^*]$ and transient eddies $[{\overline{u^\prime w^\prime}}]$ near $L_s=90^\circ$ (b, d). (a, b) and (c,d) are mass-weighted momentum fluxes in the stratosphere and troposphere (above and below $\rm p\sim 100\,mbars$), respectively. The unit for these quantities is $[m^2s^{-2}]$.}
\end{figure}

So far, we have not discussed the difficulties in spinning up stratospheric superrotation with the presence of topography. The exact cause of this issue requires detailed wave-mean analysis throughout the Titan year, which shall includes the effect of orographic waves on the zonal wind in addition to the jet acceleration due to the Rossby-Kelvin instability mechanisms suggested by previous studies \citep{Lewis2023TAMWaves, Lian2025, Wang2014}. As a demonstration of potential impact of topography on the zonal wind, we show eddy momentum fluxes using the decomposition method presented in Section~\ref{methane_transport}. Figure~\ref{fig:uw_covariance} shows the mass-weighted vertical fluxes of the zonal momentum $uw$, which represents orographic waves better than the horizontal eddy momentum fluxes. Compared to the methane fluxes (Fig.~\ref{fig:vq_covariance}), the momentum fluxes due to the eddies are more comparable to those due to the mean advection. The stationary eddy momentum flux  $[{\overline{u}}^*{\overline{w}}^*]$ is similar to the transient eddy momentum flux  $[{\overline{u^\prime w^\prime}}]$ in the troposphere but dominates in the stratosphere.  The impact of strong stationary eddies in the stratosphere is that, once dissipated due to instabilities, they can slow down the superrotation. We will present detailed analysis on the eddy-mean interactions in our Part III paper.

\section{Acknowledgment}

We thank Dr. Ralph Lorenz for many useful comments and suggestions on the simulated wind, tracer and surface/near-surface features, as well as their potential impact on the Dragonfly mission.

\section{Appendix}
\label{appendix}

Traditionally the global circulation models assume that the atmosphere is very thin compared to the radius of planet \citep{Holton2004}, as most of these models are originally intended for Earth's atmosphere. The variation of some physical and dynamical quantities in terms of the radial distance from the center of the planet is often ignored. However, depending on the region of interest, the atmosphere on some planets such as the gas giants and Titan are relatively thick compared to the radii of the planets. Using a thickness of 600 km (near the mesopause) as an example, the gravity can drop by as much as $35\%$ towards the model top.  Hereby, we make corrections to the gravitational acceleration and curvature terms (also known as metric terms) to factor in the change of radial distance from the center of Titan.

\subsection{Variable gravity}

The correction to the gravity as a function of radial distance is straightforward to implement in the MPAS dynamical core. For simplicity, we assume that the gravity is piecewise constant in each vertical layer. The definition of the variable gravity is
\begin{equation}
    g(z)=g_0 \frac{a^2}{\left(a+z\right)^2}
\end{equation}
where $g_0=\frac{GM}{a^2}$ is the gravitational acceleration at the surface of Titan, $G$ is the gravitational constant, $M$ is the mass of Titan, $a$ is the planet radius and $z$ is the distance from the surface. $g(z)$ is applied to the vertical momentum equation, as well as to the physics parameterizations where $g$ is used ({\it e.g.}, when estimating column mass via $\frac{dp}{g}$). TitanMPAS has typically 60 vertical layers that extend from the surface to about 550 km, which means the average vertical layer thickness is about 10 km. Thus, the variation of gravity in each layer is relatively small and can be approximated by the classic thin-atmosphere assumption.

\subsection{Correction to curvature terms}

In the spherical coordinates, there are several acceleration terms arise due to the spherical geometry of the planet in the momentum equations. These terms are often called curvature terms or metric terms that apply to both horizontal and vertical momentum equations. In TitanMPAS (likewise in TitanWRF), the terms need to be corrected are the Coriolis-like curvature term $\frac{uv \tan \phi}{r}$,  the centrifugal-like curvature terms $\frac{uw}{r}$ and $\frac{vw}{r}$, and the vertical curvature term $\frac{u^2 + v^2}{r}$, where $u$ is the zonal velocity, $v$ is the meridional velocity, and $w$ is the vertical velocity, $\phi$ is the latitude. The corrections are made by replacing $r=a$ with $r=a+z$. Likewise, these acceleration terms can be altered by almost $20\%$ towards the model top.

\bibliography{bibliography}{}
\bibliographystyle{aasjournal}

\end{document}